\documentclass[journal]{IEEEtran}

\usepackage{amsmath}
\usepackage{graphicx}
\usepackage{siunitx}

\usepackage{xcolor}
\usepackage{listings}
\usepackage{makecell}

\usepackage{xparse}

\NewDocumentCommand{\codeword}{v}{%
\texttt{\textcolor{blue}{#1}}%
}
\usepackage{xcolor}
\definecolor{light-gray}{gray}{0.95}

\usepackage{hyperref}
\ifCLASSINFOpdf
\else
\fi

\begin{document}

\title{Mitigating Mismatch Compression in Differential Local Field Potentials}

\author{Vineet~Tiruvadi,
        Sam~James,
        Bryan~Howell,
        Mosadoluwa~Obatusin,
        Andrea~Crowell,
        Patricio~Riva-Posse,
        Ki~Sueng~Choi,
        Allison~Waters,
        Robert~Gross,
        Cameron~C.~McIntyre,
        Helen~Mayberg,
        and~Robert~Butera
\thanks{V. Tiruvadi is with Department of Biomedical Engineering and School of Electrical and Computer Engineering, Georgia Institute of Technology and Emory School of Medicine, Atlanta, GA, USA}
\thanks{S. James was with School of Electrical and Computer Engineering, Georgia Institute of Technology, Atlanta, GA, USA}
\thanks{B. Howell and C. McIntyre are with Duke University, Durham, NC, USA}
\thanks{M. Obatusin, H. Mayberg, K.S. Choi, A. Waters are with Icahn School of Medicine at Mt. Sinai, New York, NY, USA}
\thanks{A. Crowell, and P. Riva-Posse are with Emory School of Medicine, Atlanta, GA, USA}
\thanks{R. Gross is with Department of Biomedical Engineering, Georgia Institute of Technology, and Emory School of Medicine, Atlanta, GA, USA}
\thanks{R. Butera is with Department of Biomedical Engineering, Georgia Institute of Technology, Atlanta, GA, USA}
}

\markboth{IEEE Transactions on Neural Systems and Rehab Eng X}%
{Tiruvadi \MakeLowercase{\textit{et al.}}: Mitigating Mismatch Compression in Differential Local Field Potentials}

\maketitle

\begin{abstract}
    Bidirectional deep brain stimulation (bdDBS) devices capable of recording differential local field potentials ($\partial$LFP) enable neural recordings alongside clinical therapy.
    Efforts to identify objective signals of various brain disorders, or \textit{disease readouts}, are challenging in $\partial$LFP, especially during active DBS.
    In this report we identified, characterized, and mitigated a major source of distortion in $\partial$LFP that we introduce as \textit{mismatch compression} (MC).
    MC occurs secondary to impedance mismatches across the $\partial$LFP channel resulting in incomplete rejection of artifacts and downstream amplifier gain compression.
    Using \textit{in silico} and \textit{in vitro} models we demonstrate that MC accounts for impedance-related distortions sensitive to DBS amplitude.
    We then use these models to develop and validate a mitigation strategy for MC that is provided as an opensource library for more reliable oscillatory disease readouts.
\end{abstract}


\IEEEpeerreviewmaketitle
\section{Introduction}
    Deep brain stimulation (DBS) is an increasingly effective therapy for neuropsychiatric disorders \cite{perlmutter2006,mayberg2005,holtzheimer2011}.
    Efforts to automate DBS adjustments using neural recordings are growing through adaptive algorithms that use physiologically-derived \textit{disease readouts} \cite{widge2018closing,swann2018adaptive}.
    A new generation of bidirectional DBS devices \cite{starr2018totally,anso2022concurrent} enable direct recording of local field potential (LFPs) from patients receiving DBS therapy, with \textit{differential} LFP ($\partial$LFP) recordings emerging as a popular strategy \cite{gilron2021,starr2018totally,stanslaski2012,stanslaski2018}.
    While modern efforts are making significant progress in ensuring these LFPs accurately reflect, care must be taken to account for artifacts that are intrinsic to \textit{differential} recording setups \cite{anso2022concurrent,stanslaski2018}.
    
    $\partial$LFP channels record from two electrodes to remove large-voltage artifacts recorded equally in both electrodes before reaching sensitive recording hardware \cite{stanslaski2012, meyer2018differential}.
    However, heterogeneous brain targets can prevent the stimulation artifact from being equally recorded by both electrodes, resulting in incomplete rejection.
    Even a relatively small unrejected artifact can then overwhelm downstream amplifiers tuned for much smaller signals, a failure mode of signal amplifiers well known as \textit{gain compression} \cite{stanslaski2012}.
    Overt gain compression, appearing as a visually identifiable clipping of the recording, is often identified visually, but realistic amplifiers can fail gradually and introduce the same frequency-domain distortions in a more subtle form.
    This process, which we introduce as \textit{mismatch compression} (MC), distorts downstream oscillatory analyses and confounds disease readouts with impedance measurements.
    While next-generation DBS devices have reduced sources of noise, MC distortions are intrinsic to $\partial$LFP channels are must be ruled out or accounted for in chronic readouts \cite{stanslaski2012, cummins2021chronic}.
    Simulations, in particular, are useful tools to determine whether putative signals in recordings are of neural or artifactual sources.
    
    Here, we demonstrated, characterized, and mitigated MC distortions in oscillatory power analyses.
    Clinical recordings were measured as a part of a study of subcallosal cingulate white matter (SCCwm) DBS for treatment resistant depression (TRD), and used as exemplars of potentially distorted recordings.
    We observed distortions consistent with gain compression in \textit{in vivo} clinical recordings then confirmed and characterized MC using a \textit{in silico} reduced model and \textit{in vitro} benchtop models of $\partial$LFP.
    We then use these models to design and validate an opensource MC mitigation strategy that can enable more reliable oscillatory analyses.

    \begin{figure}[!ht]
        \centering
        \includegraphics{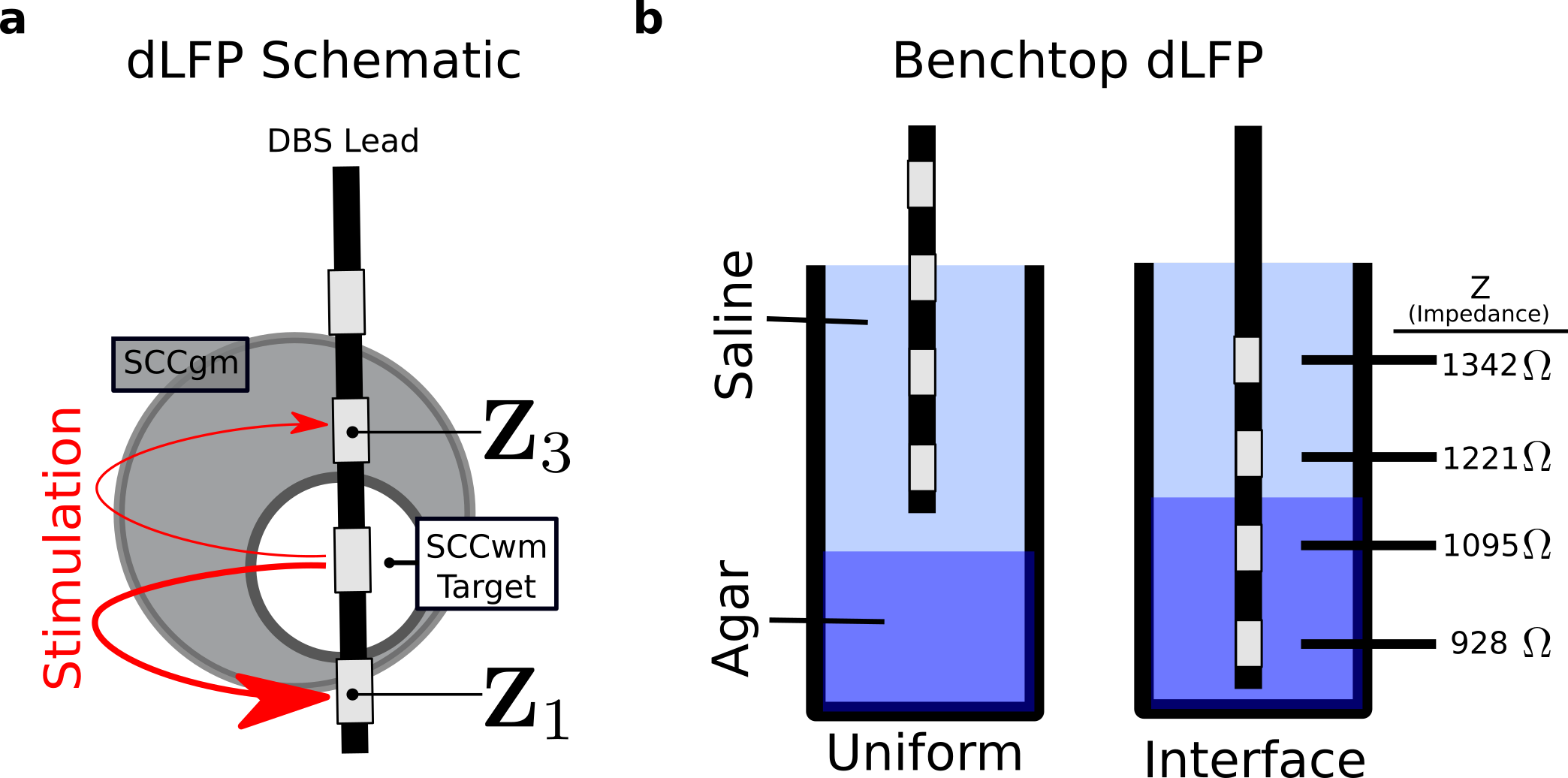}
        \caption{\textbf{Schematic of Differential Local Field Potential}. a) Electrodes in non-uniform brain tissue exhibit different impedances. The resulting impedance mismatch $Z_\Delta$ leads to different amplitudes of recorded stimulation and imperfect common-mode rejection, a process called \textit{mismatch compression}. b) \textit{In vitro} testing using an agar construct with variable resistivity was used to confirm and address mismatch compression. Two phases (light blue, dark blue) are constructed with variable impedances.}
        \label{fig:dLFP_overview}
    \end{figure}
    
\section{Methods}
    \subsection{Clinical Protocol} 
        Six patients with treatment resistant depression (TRD) were consented and enrolled in an IRB and FDA approved research protocol at Emory University (clinicaltrials.gov NCT01984710; FDA IDE G130107) (Table \ref{tab:pts}).
        Two Medtronic DBS3387 leads (Medtronic PLC, Minnesota, USA) were implanted bilaterally in patient-specific, tractography-defined subcallosal cingulate white matter (SCCwm) as previously described \cite{riva2018connectomic} (Figure \ref{fig:dLFP_overview}a).
        Each lead has four electrodes with edge-to-edge spacing \SI{1.5}{\milli\meter} and was connected to a prototype Activa PC+S\texttrademark\space (Medtronic PLC, Minnesota, USA) implantable pulse generator (IPG).
        
    \subsection{Stimulation and Impedance}
        Therapeutic DBS stimulation is delivered bilaterally at \SI{130}{\hertz} in monopolar mode with the IPG as cathode.
        All electrode impedances were measured in monopolar mode at \SI{3}{\volt} and \SI{100}{\hertz} using the standard clinician-controller (Medtronic N'Vision) \cite{kent2015}.
        Impedance mismatch was calculated by subtracting the impedances of the two recording electrodes at each weekly visit.
        Measurements were made at weekly clinical patient assessments over 28 weeks post-implantation.
        The same procedure was used to measure impedances in \textit{in vitro} lead experiments.
        
    \subsection{Recording Parameters}
        \begin{table}
		\centering
		\begin{tabular}{c | c | c | c | c |} 
			\thead{Patient} & \thead{Age} & \thead{Sex} & \thead{Recording Electrodes} & \thead{Gains}\\
			\hline
			Patient 1 & 50 & F & (E1,E3)+(E8,E10) & 250,2000\\
            Patient 2 & 48 & F & (E1,E3)+(E9,E11) & 1000,1000\\
			Patient 3 & 70 & F &  (E1,E3)+(E8,E10) & 2000,2000\\
			Patient 4 & 64 & M & (E1,E3)+(E9,E11) & 2000,2000\\
			Patient 5 & 62 & F & (E0,E2)+(E8,E10) & 250,250\\
			Patient 6 & 57 & M & (E1,E3)+(E8,E10) & 250,250 \\	
		\end{tabular}
		\caption{\textbf{Patient Demographics and Parameters}. Six patients were included in this study. Recording electrodes are chosen around the ultimate therapeutic contact. Gains are set for (Left, Right) channels independently after visual inspection of recorded LFP.}
		\label{tab:pts}
	    \end{table}
	    
        All PC+S\texttrademark recordings were sampled at \SI{422}{\hertz}, constrained by the device and recording capacity.
        Hardware filters were set at \SI{0.5}{\hertz} high-pass and \SI{100}{\hertz} low-pass.
        Each channel has an adjustable gain parameter, selected from 200,400,1000,or 2000, which is set in patients after visual inspection of recording spectrograms by a Medtronic device engineer.
        A constant amplitude \textit{over-range marker} (ORM) at \SI{105.5}{\hertz} is a part of all PC+S\texttrademark\space recordings and is modulated in the setting of overt amplifier saturation.
        \textbf{\textit{In vivo}} recordings were taken from bilateral SCC using patient-specific parameters for recording electrode number and channel gains (Table \ref{tab:pts}).
        A channel consisted of two recording electrodes per lead, with the left lead labeled E0-3 and the right lead labeled E8-E11 (Figure \ref{fig:dLFP_overview}a), and chosen to be around the patient-specific optimal stimulation electrode.
        \textbf{\textit{In vitro}}, a single DBS 3387 lead was connected to the channels 0-7 on the Activa PC+S\texttrademark. 
        All recordings were taken at 1000 gain for consistency.
        \textit{In vitro} recordings were collected using the standard clinical sensing tablet and with all other parameters identical to \textit{in vivo} recordings.
        Leads were placed in either uniform resistivity or at the interface of two different resistivities, implemented as the \textit{in vitro} construct.
        
    \subsection{\textit{In vitro} Construct}
        An agar-saline preparation of two spatially sequestered phases with distinct resistivities was constructed based on \cite{kandadai2012}.
        The saline phase was fixed at \SI{0.5}{\frac{\milli\gram}{\milli\liter}} of NaCl, and yielded measured impedances of approximately \SI{800}{\ohm}.
        The agar phase was fixed with high resistivity \SI{0.1}{\frac{\milli\gram}{\milli\liter}} of NaCl), and yielded measured impedances of approximately \SI{1300}{\ohm}.
        Agar mixture was poured into a \SI{10}{\milli\liter} conical corning tube with blue fluorophore and placed in a \SI{32}{C} for 20 minutes to settle.
        Saline phase was then added on top of the settled agar phase.
        A demo DBS3387 lead is then placed at the interface of the saline and agar layers using a micromanipulator.
        Impedance mismatches measured across two non-adjacent electrodes ranged from \SI{100}{\ohm} in uniform media and \SI{300}{\ohm} across media.
        
    \subsection{\textit{In silico} Model}
        \label{sec:meth_model}
        \begin{figure}[!ht]
            \centering
            \includegraphics{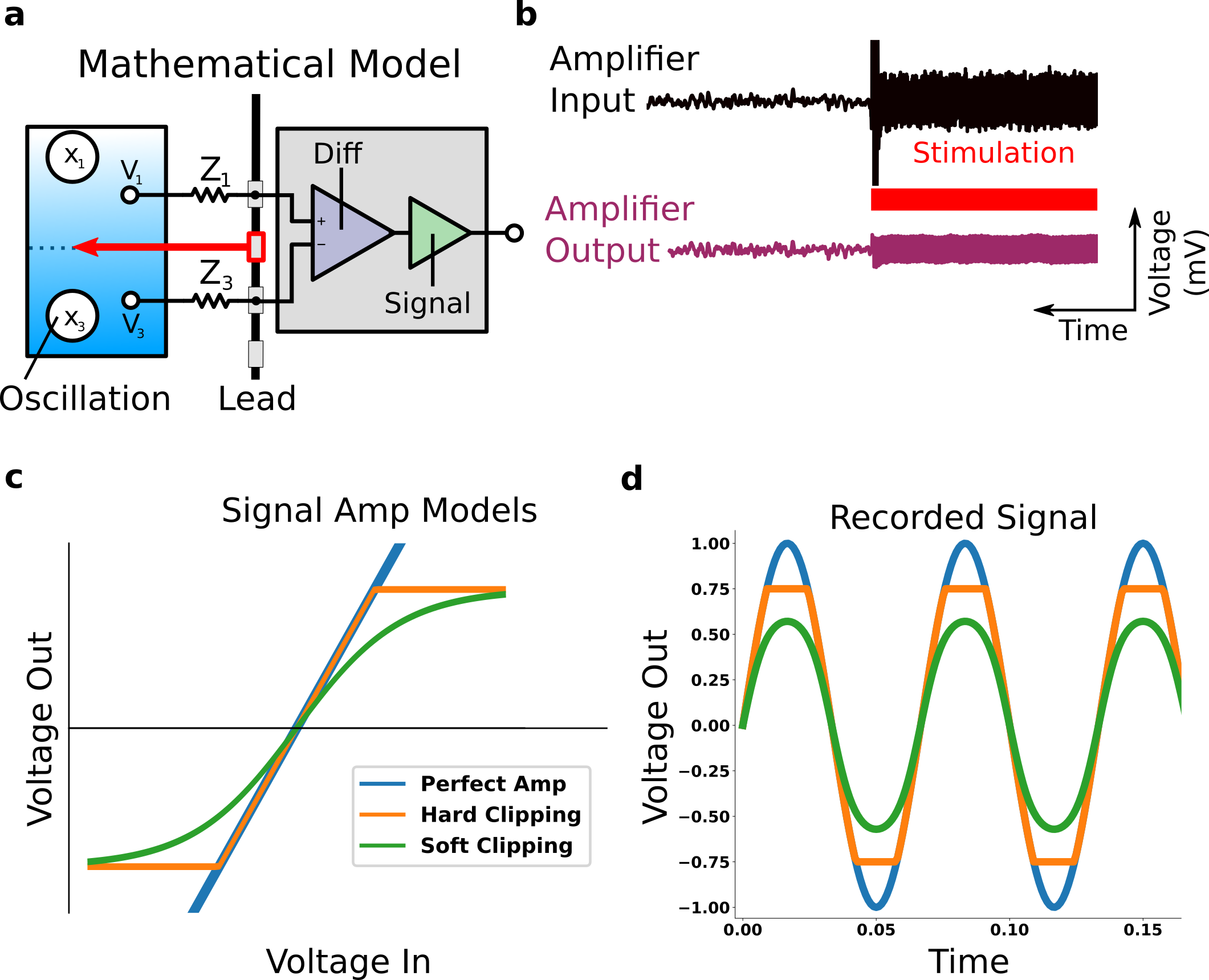}
            \caption{\textbf{Mathematical model of $\partial$LFP for mismatch compression}. \textbf{a,} Schematic of model components in a differential LFP ($\partial$LFP) recording bidirectional DBS (bdDBS) device. A neural oscillation ($x_1$) is recorded ($V_1$) at one electrode with impedance $Z_1$. Stimulation (red arrow) is introduced. The recording circuitry is split into two components: a differential amplifier and a signal amplifier. \textbf{b,} simulated inputs and associated output at various impedance mismatches. \textbf{c,} Transfer functions for three signal amplifier models: perfect linear amplifier, piecewise-linear hard-clipping amplifier, and a hyperbolic-tan soft-clipping amplifier. \textbf{d,} Effect of different models on sinusoidal output signal. Hard-clipping (orange) and soft-clipping (green) reflect two types of gain compression.}
            \label{fig:model_overview}
        \end{figure}
        
        A generic bidirectional DBS device with $\partial$LFP recording channel was modeled (Figure \ref{fig:model_overview}a).
        A single neural oscillation $x_1$ was implemented as a stationary \SI{15}{\hertz} sinusoid (Figure \ref{fig:model_overview}a), and an independent $\frac{1}{f}$ components added to $x_1$ and $x_3$ sources.
        Stimulation artifact was introduced as a truncated Fourier series of sine waves at the therapeutic stimulation $f_{T}$, yielding the \textit{stimulation shaping harmonics} (SSH).
        The stimulation waveform $S(t)$ is set six orders of magnitude larger than the neural oscillations.
        The Activa PC+S\texttrademark\space specific ORM is added at \SI{105.5}{\hertz} to improve congruence between simulation and empirical recordings.
        
        Recordings are sampled at \SI{422}{\hertz}, the dual-channel sampling rate of the PC+S\texttrademark, yielding \textit{aliased shaping harmonics} (ASH) for SSH above Nyquist rate (\SI{211}{\hertz}).
        Electrodes $e_1$ and $e_3$ measure neural sources and the stimulation artifact through independent impedances $Z_1$ and $Z_3$.
        The differential amplifier outputs the unity-gain differential mode $V_\text{diff} = e_1 - e_3$:
        \begin{equation}
        	V_\text{out} \approx A_d Z_b S(t) \left( \frac{1}{Z_1 + Z_b} - \frac{1}{Z_3+Z_b} \right)
        	\label{eq:diff_v_out}
        \end{equation}
        This differential mode then goes to the signal amplifier, implemented as one of three transfer functions: linear corresponding to perfect linear, hard-clipping corresponding to piecewise linear, and soft-clipping corresponding to $\tanh$ (Figure \ref{fig:model_overview}c).
        Soft-clipping (Figure \ref{fig:model_overview}d) yields a waveform that is gain compressed but can be difficult to identify in the time-domain.
        \begin{equation}
            V_\text{lfp} = g_2 \cdot \tanh(g_1 \cdot V_\text{out})
        \end{equation}
        Each model distorts the SSH+ASH to different degrees, yielding \textit{intermodulation harmonics} (IMH).
        The output of the signal amplifier was then considered the $\partial$LFP recording and analysed.
        
    \subsection{Model Parameters}
        \begin{table}
			\centering
			\begin{tabular}{c c c c c c} 
				$A_d$ & $Z_{b}$ & $\frac{1}{f}$ Strength & $x_1$ amplitude & $g_1$ & $g_2$ \\ \hline
				$500$ & $1\cdot 10^{4} \Omega$ & $1\cdot10^{-3}$ & $2 \cdot 10^{-3}$ & 1 & 1 
			\end{tabular}
			\caption{Model parameters used for \textit{in silico} simulated recordings. $A_d$ is amplitude of the stimulation artifact after common-mode rejection. $Z_b$ is the internal impedance of the differential channel. $x_1$ amplitude is the strength of the oscillation. $g_1$ is the post-differential amplifier gain, $g_2$ is the post-signal amplifier gain.}
			\label{tab:A1_ModelParams}
		\end{table}
        Model parameters (Table \ref{tab:A1_ModelParams}) were chosen to yield simulated $\partial$LFP with visual similarity to empirical, specifically with respect to broad spectrum shape.
        More precise, systematic fitting was avoided due to the presence of numerous other artifacts and our specific focus on MC as a mechanism \cite{stanslaski2012,swann2017chronic}.
        
    \subsection{Oscillatory Analyses}
        $\partial$LFP were analysed in the frequency domain using a Welch power spectral density (PSD) estimate with 1024 FFT bins, 0\% overlap, 844 sample Blackman-Harris Window.
        PSDs were log-transformed $10 \cdot \log_{10}(P_{xx})$ to visualize logPSD and perform preprocessing.
        Oscillatory power was then computed as either the mean or median value of the PSD for a predefined frequency range corresponding to standard oscillatory bands: $\delta$ (\SIrange{1}{4}{\hertz}), $\theta$ (\SIrange{4}{8}{\hertz}), $\alpha$ (\SIrange{8}{14}{\hertz}), $\beta$ (\SIrange{14}{30}{\hertz}), $\gamma$ (\SIrange{30}{50}{\hertz}) \cite{veerakumar2019field}.
        
    \subsection{Analysis and Simulation Code}
        Analyses and simulation were done through open-source Jupyter Notebooks available at \url{https://github.com/virati/mismatch_compression}.
        Dependencies and associated libraries are available through PyPi: NumPy\cite{harris2020array}, SciPy\cite{virtanen2020}, Allantools\cite{wallin_allantools}, and DBSpace \cite{dbspace}.
        
\section{Results}
    \subsection{Clinical \(\partial\)LFP demonstrate significant variability}
        \begin{figure}
            \centering
            \includegraphics{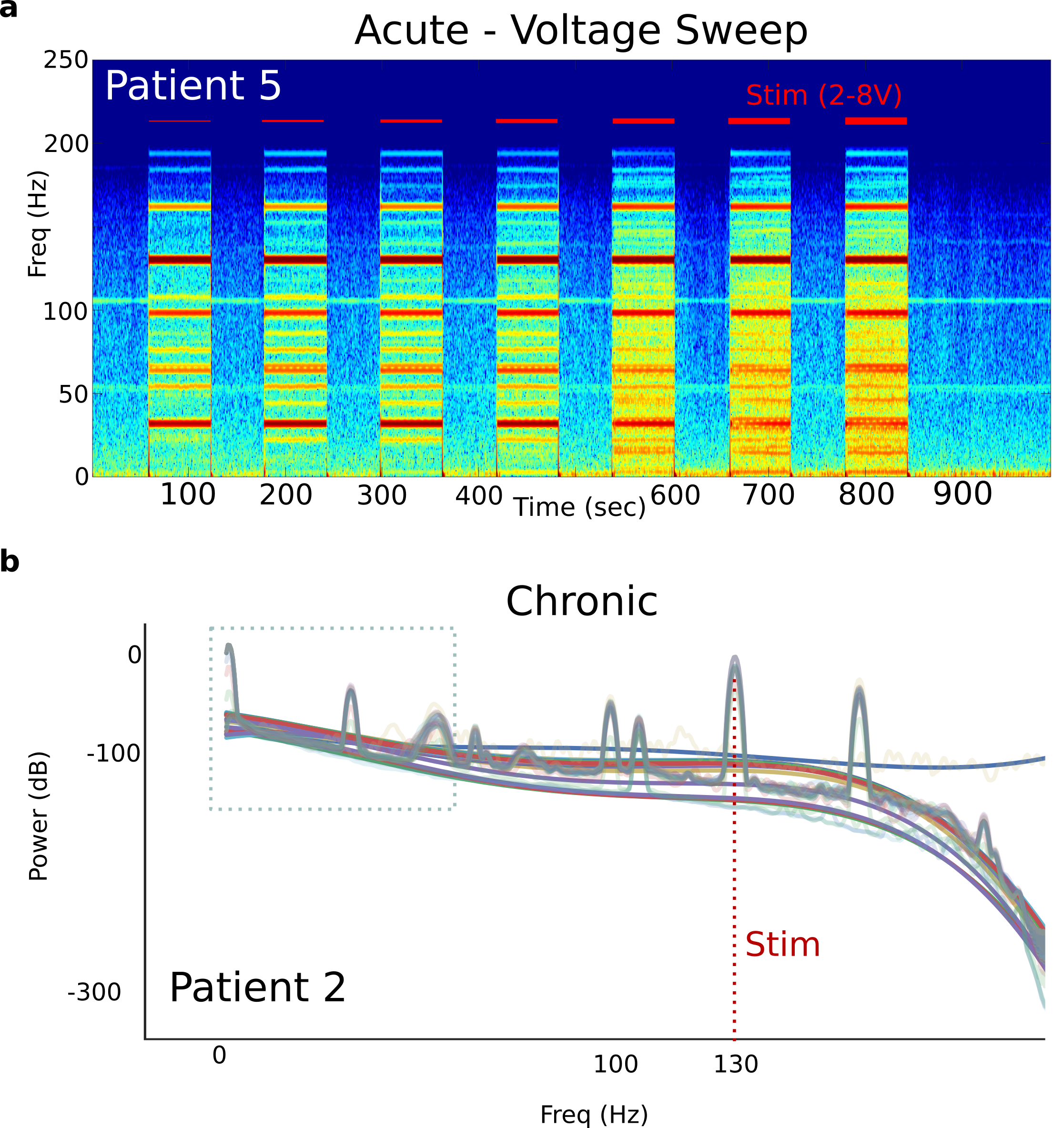}
            \caption{\textbf{Significant variability evident in recordings}. \textbf{a,} Experiments involving stimulation at variable amplitudes demonstrate stimulation-locked changes in frequency content. The nature of these changes are not consistent with typical neural sources. \textbf{b,} PSDs from $\partial$LFP recordings in a patient demonstrate significant variability across months of recording. Each color curve (translucent) corresponds to a different weekly averaged PSD. A fourth order polynomial is fit to each average PSD (solid lines) to visualize variability in broad-spectrum features.}
            \label{fig:in_vivos}
        \end{figure} 
        
        \textit{In vivo} clinical recordings from patients exhibit large differences between no DBS and active DBS.
        Recordings taken during experiments with active stimulation change significantly as a function of the stimulation voltage (Figure \ref{fig:in_vivos}a).
        Over months, recording PSDs exhibit variability, both within a patient (Figure \ref{fig:in_vivos}b) and across patients (data not shown).
        While a neural source for these large changes was feasible, the sharp features of the PSDs during stimulation suggested stimulation artifact unless otherwise proven.
        
    \subsection{$\partial$LFP recording environment is dynamic}
        \begin{figure}[!ht]
            \centering
            \includegraphics{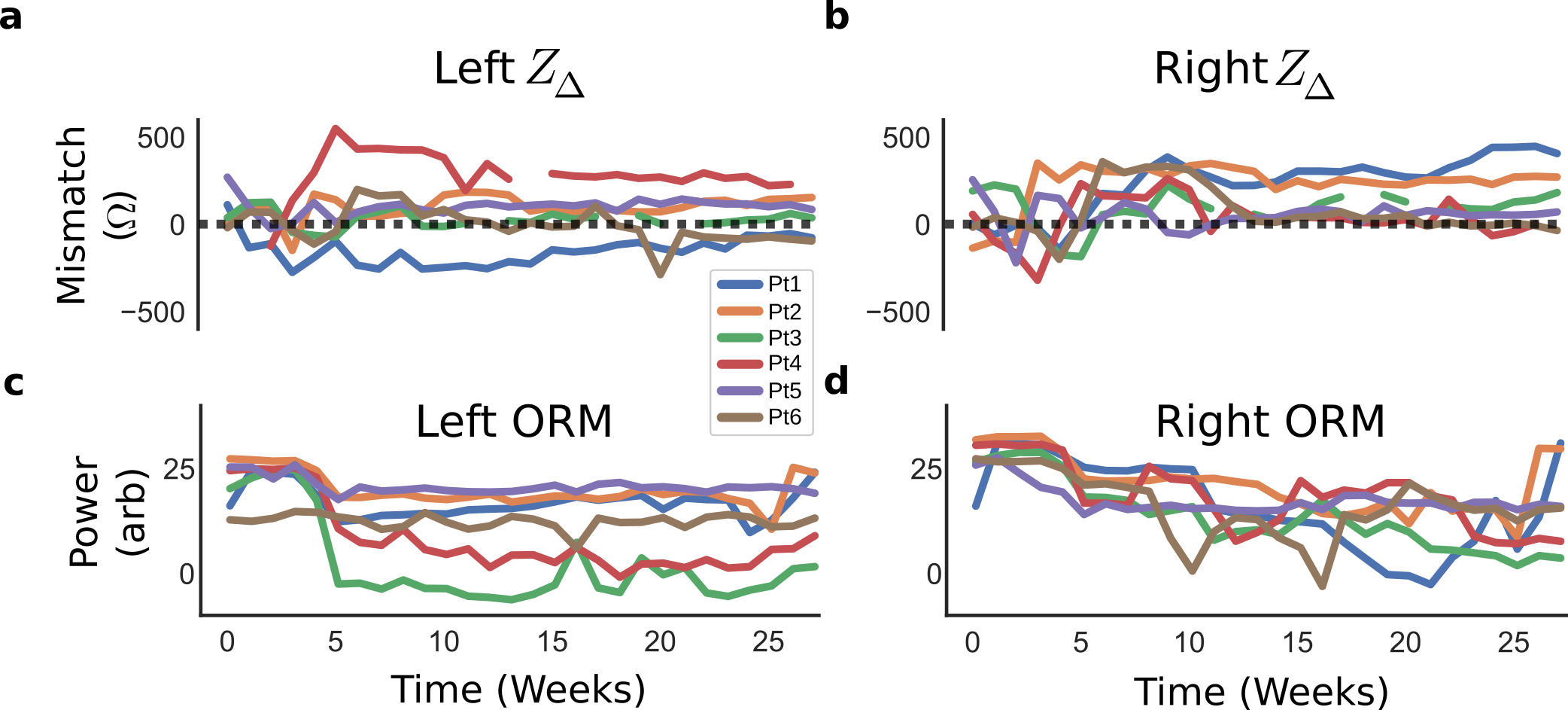}
            \caption{\textbf{$\partial$LFP channel instabilities}. Impedances in the recording electrodes of all patients demonstrate significant, dynamic mismatches in both a) left and b) right DBS leads. c,d) Mismatches between the two recordings electrodes were calculated and demonstrated significant variance between patients, over time, and between hemispheres. e,f) Over-range marker (ORM) was calculated across recordings from all weeks to assess whether saturation is potentially occurring.}
            \label{fig:Z_diff}
        \end{figure}
        Weekly impedance mismatches ranged between \SI{0}{\ohm} and \SI{600}{\ohm} with significant variability across time, between patients, and between leads (Figure\ref{fig:Z_diff}a,b).
        We calculated power in the ORM and observed large variation between patients and over the study timecourse (Figure \ref{fig:Z_diff}c,d).
        Left and right channel ORM change differently from each other.
        
        In both impedance mismatch and ORM measurements, two distinct phases were consistently seen across patients: a variable phase (0-10 weeks) and a stable phase (11-28 weeks), with significant variability between patients through all phases.
        
    \subsection{Mismatch compression regenerates distortions}
        \begin{figure*}[!ht]
            \centerline{\includegraphics{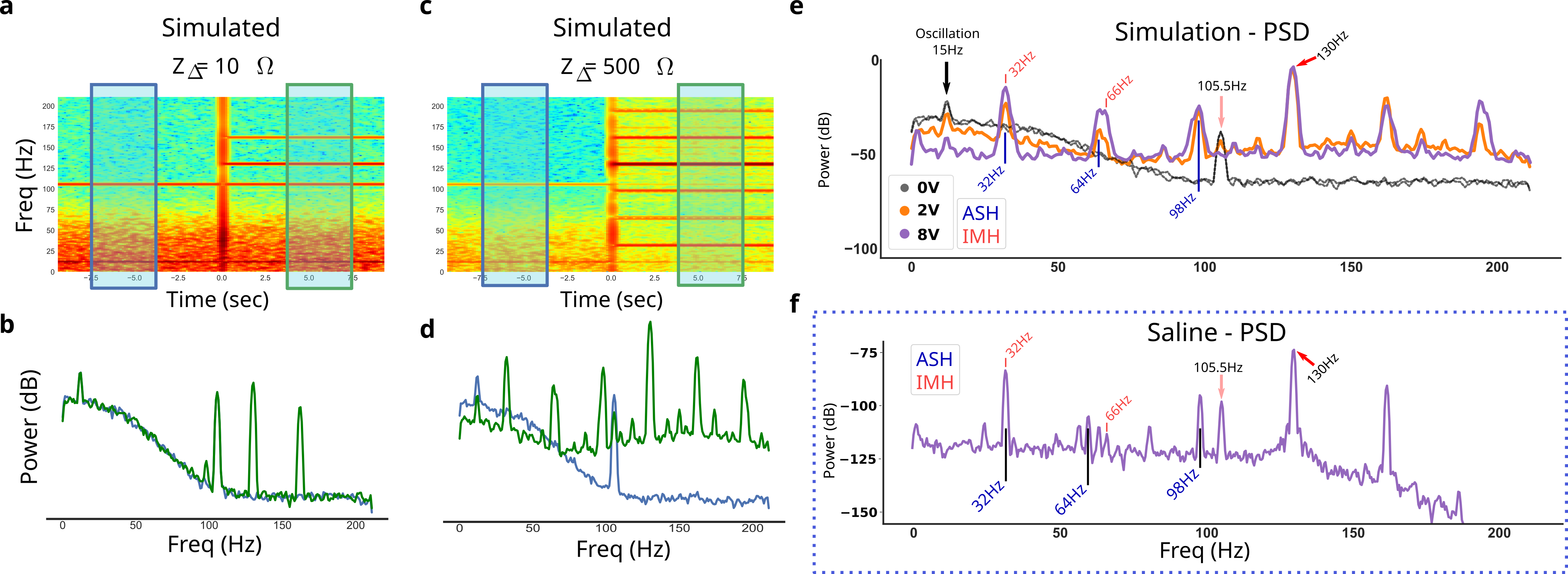}}
            \caption{\textbf{Simulation regenerates large empirical peaks}. \textbf{a,b,} Simulated $\partial$LFP at a low impedance mismatch ($Z_\Delta$) shows artifacts during simulated \SI{130}{\hertz} stimulation. Before stimulation (blue box and line) is compared to during stimulation (green box and line). \textbf{c,d,} Simulated $\partial$LFP at a high impedance mismatch shows significantly more artifactual peaks, as well as broad slope changes and low-frequency distortions. \textbf{e,} Simulation at various voltages shows the effect of gain compression and artifact locations. Aliased simulation harmonics (ASH) and intermodulation harmonics (IMH) are labeled. \textbf{f,} Empirical recording in saline shows a peak at all simulated peaks. Additional peaks are evident and are ascribed to other device-related artifacts.}
            \label{fig:stim_arts}
        \end{figure*}
        We hypothesized \textit{mismatch compression} (MC) as a major contributor to measurement given the design rationale outlined in \cite{stanslaski2012}.
        To test this hypothesis we used a $\partial$LFP model of MC to simulate recordings at various impedance mismatches (Figure \ref{fig:stim_arts}a,b).
        Grossly, the MC model was able to simulate broad slope changes and the emergence of IMH solely by adjusting the impedance mismatch (Figure \ref{fig:stim_arts}b,d).
        
        Measurement in uniform saline are consistent with the predicted ASH and IMH artifacts (Figure \ref{fig:stim_arts}e,f).
        In particular, the distinction between the SAH \SI{64}{\hertz} artifact and the IMH \SI{66}{\hertz} is evident as a broader multipeak artifact, present in both \textit{in silico} (Figure \ref{fig:stim_arts}e) and \textit{in vitro} models (Figure \ref{fig:stim_arts}f).
        Interestingly, MC induces an absolute reduction of the simulated constant \SI{15}{\hertz} neural oscillation as a function of stimulation voltage (Figure\ref{fig:agar_exp}a).
        Other artifacts are evident \textit{in vivo} that are not generated by the MC model (Figure \ref{fig:stim_arts}f).
        
    \subsection{\textit{In vitro} resistivity mismatches distort $\partial$LFP}
        \begin{figure*}[!ht]
            \centering
            \includegraphics{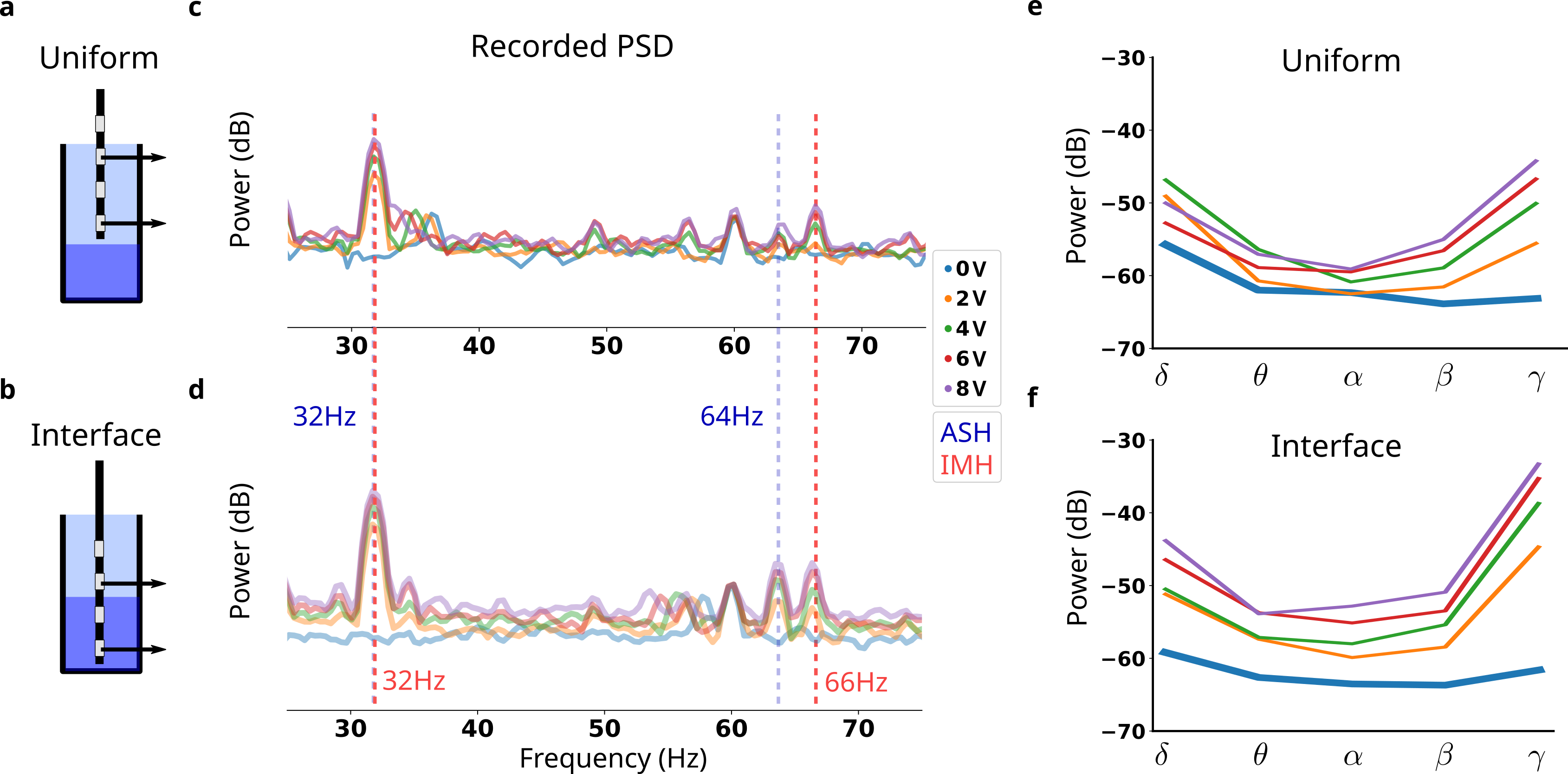}
            \caption{\textbf{\textit{In vitro} mismatch compression}. $\partial$LFP recordings were captured in two configurations: \textbf{a,} uniform saline medium and \textbf{b,} interface of saline-agar. \textbf{c,} Uniform-medium PSDs at various voltages demonstrate distinct peaks, with \SI{32}{\hertz}, \SI{64}{\hertz}, and \SI{66}{\hertz} being highlighted for their voltage-dependence. \textbf{d,} Interface-media PSDs demonstrate more voltage-dependence, both in the highlighted peaks and in broad-spectrum slope. \textbf{e,} Oscillatory power calculated in uniform-medium at various stimulation voltages compared to \textbf{f,} interface-medium.}
            \label{fig:agar_exp}
        \end{figure*}
        To experimentally verify the MC hypothesis, \textit{in vitro} $\partial$LFP recordings are measured at different impedance mismatches.
        Measurements in uniform media demonstrates emergence of IMH, particularly the \SI{66}{\hertz} peak (Figure \ref{fig:agar_exp}a,c).
        Measurement at the interface of saline and agar interface demonstrated larger artifacts.
        The level of MC distortion is sensitive to the resistivity mismatch across the $\partial$LFP channel (Figure \ref{fig:agar_exp}c,d).
        \textit{In vitro} $\partial$LFP recordings using the PC+S\texttrademark\space confirmed that distortions depended on the mismatch in resistivity of the two recording electrodes (Figure \ref{fig:agar_exp}) and that this mismatch resulted in predicted artifacts (Figure \ref{fig:stim_arts}).
        
        Broadband flattening of the PSD slope is observed with increasing stimulation voltage, more vividly in the interface recordings than the uniform recordings (Figure \ref{fig:agar_exp}d).
        Oscillatory powers calculated in each \textit{in vitro} mismatch condition reflected the level of MC, with interface recordings having broadly higher measured power in all bands (Figure \ref{fig:agar_exp}e,f).
    
    \subsection{Assessing Gain Compression Ratio (GCr)}
        \begin{figure}
            \includegraphics{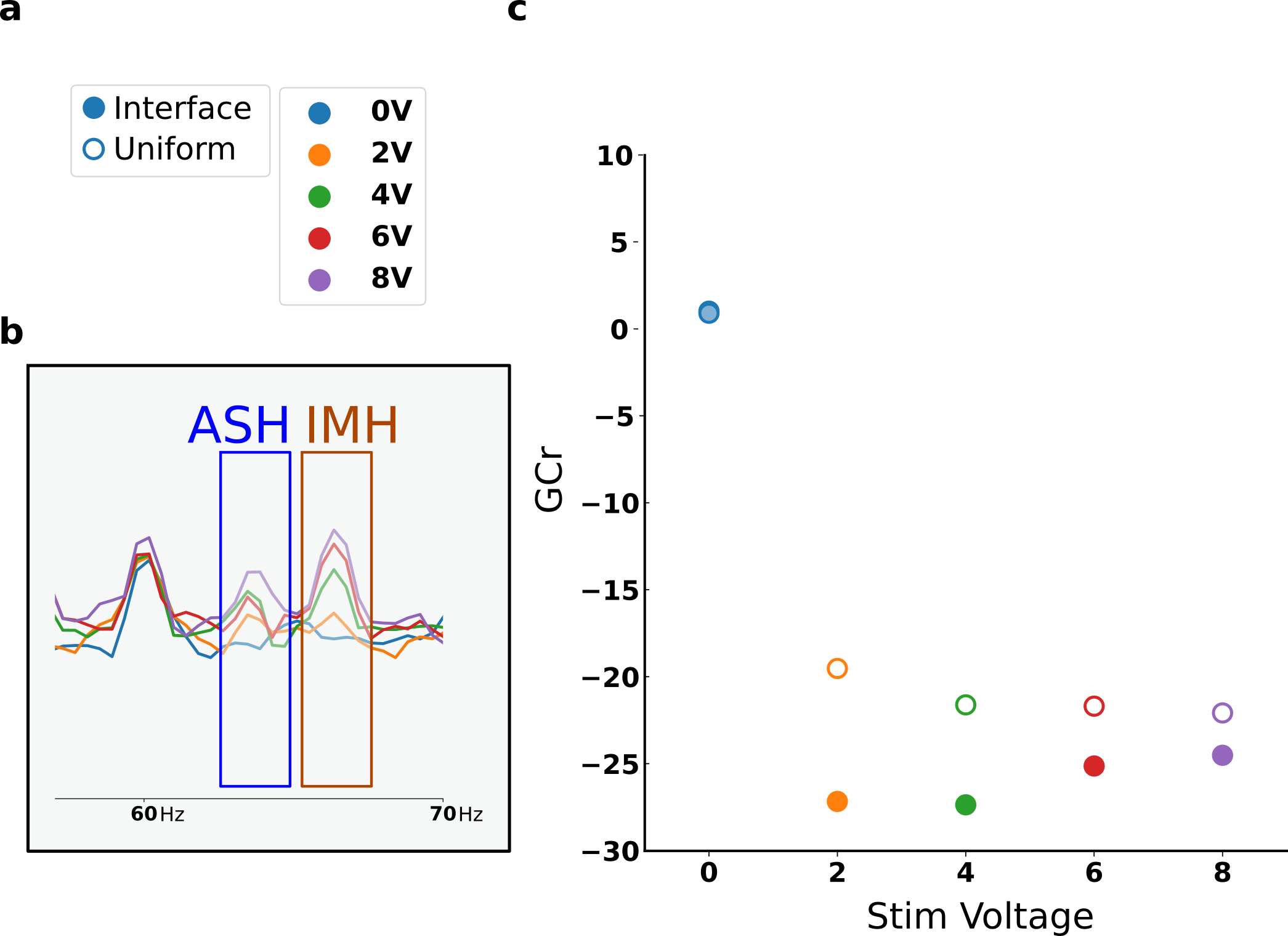}
            \caption{\textbf{MC distortions}. \textbf{a,} Recordings taken at both interface (solid dot) and uniform (empty dot), with a range of stimulation voltages \SIrange{0}{8}{\volt}. \textbf{b,} Aliased stimulation harmonic (ASH) arises from stimulation shaping harmonics (SSH) suboptimally sampled, while intermodulation hamornics (IMH) arise directly from amplifier gain compression. \textbf{c,} Gain compression ratio (GCr) calculated from ASH and IMH reflect both stimulation voltage and impedance mismatch.}
            \label{fig:gcr}
        \end{figure}
        Comparing the power in the ASH artifacts (\SI{64}{\hertz}) versus the IMH (\SI{66}{\hertz}) (Figure \ref{fig:gcr}) can reflect the level of gain compression.
        Stimulation voltages are swept between \SIrange{0}{8}{\volt} in \SI{1}{\volt} steps, at both uniform and interface surroundings (Figure \ref{fig:gcr}a).
        To assess the level of gain compression occurring, a gain compression ratio (GCr) is calculated ASH+IMH and IMH (Figure \ref{fig:gcr}b).
        The GCr is calculated under each stimulation voltage at both mismatch conditions (Figure \ref{fig:gcr}c).
        \SI{2}{\volt} in the interface mismatch condition evokes a similar GCr as the \SI{8}{\volt} in the uniform mismatch condition.
        
    \subsection{Mitigating mismatch compression}
        \label{sec:mc_mitigation}
        \begin{figure*}[!ht]
            \centering
            \includegraphics{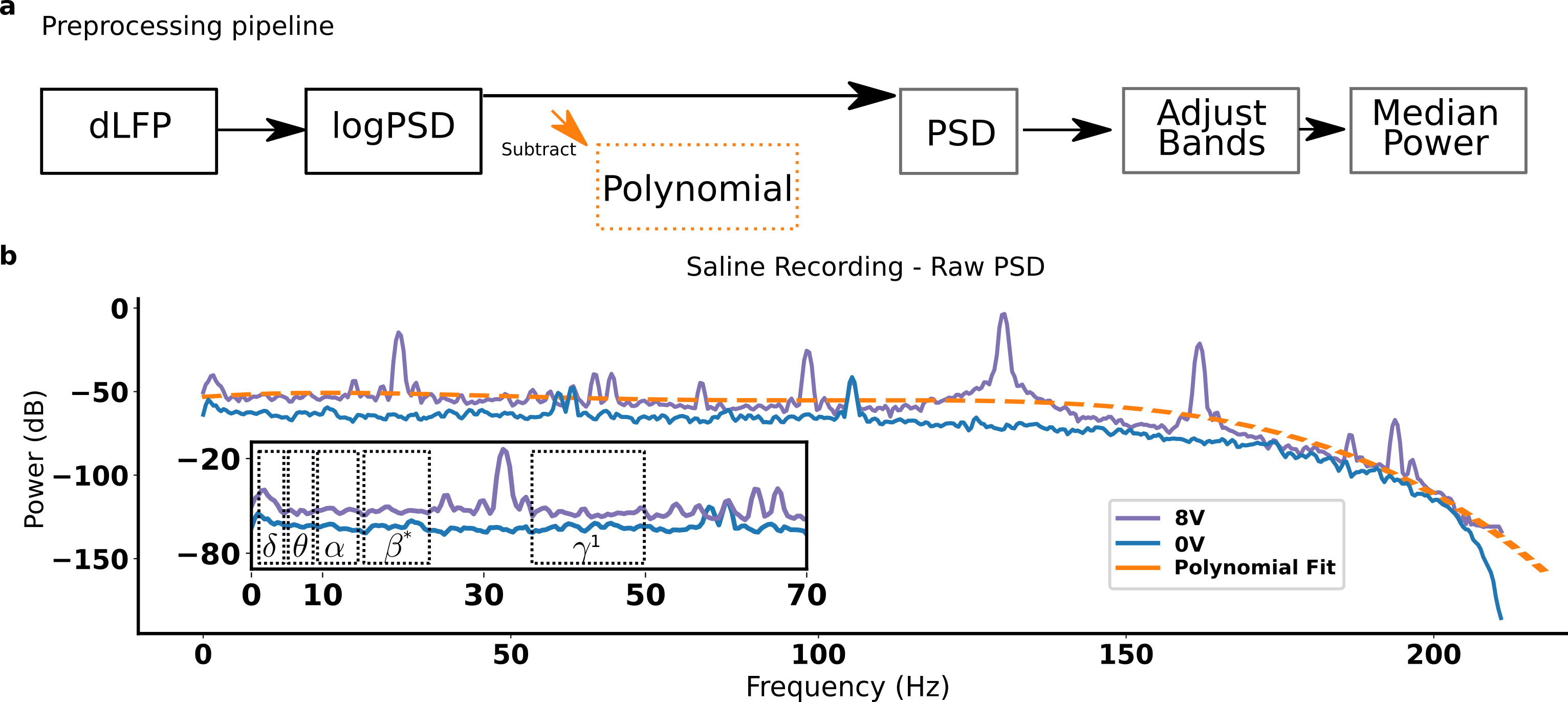}
            \caption{\textbf{MC mitigation pipeline} - \textbf{a,} MC mitigation pipeline for $\partial$LFP recordings remove frequency features that can be distorted. \textbf{b,} Empirical PSD in saline at two stimulation voltages demonstrates mismatch compression. Inset- Adjustments to the frequency windows for oscillatory bands to avoid mismatch compression artifacts.}
            \label{fig:mc_mitigation}
        \end{figure*}
\begin{figure*}
            \centering
            \includegraphics{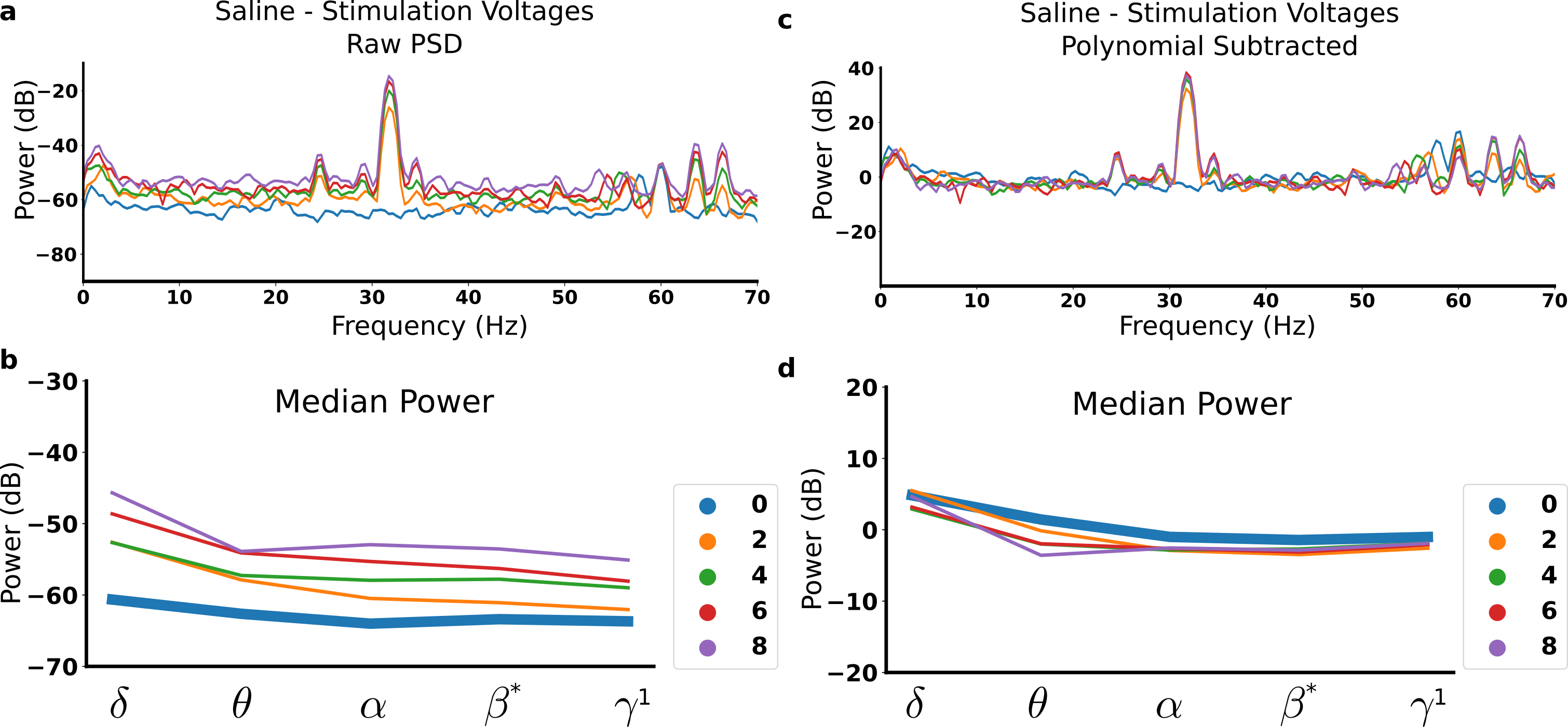}
            \caption{\textbf{MC corrections \textit{in vitro}}. \textbf{a,} Raw PSDs recorded in agar at stimulation voltages between \SIrange{0}{8}{\volt}. \textbf{b,} Oscillatory power calculated in each band for each tested stimulation voltage. \textbf{c,} Corrected PSDs remove features that are corrupted by mismatch compression. \textbf{d,} Oscillatory power calculated converges to the no-stimulation condition across all stimulation voltages.} 
            \label{fig:PSD_correct}
        \end{figure*}

        To mitigate MC, we propose removing features susceptible to distortion.
        Using the reduced \textit{in silico} model we identified three simple preprocessing steps.
        First, polynomial subtraction removes broadband MC distortions that flatten the PSD slope \ref{fig:mc_mitigation}b).
        The order of the polynomial fit was based from prior literature \cite{wang2016} and fixed at a fourth-order polynomial (Figure \ref{fig:mc_mitigation}b, dotted orange line).
        Second, median power within adjusted oscillatory bands (Table \ref{tab:prelim_oscbands}) are more robust to distortions (Figure \ref{fig:mc_mitigation}b, inset).
        Band ranges were chosen using both \textit{in silico} and \textit{in vitro} LFP during stimulation to find the maximal continuous range in standard oscillatory bands that also avoided ASH+IMH.
        Finally, a gain compression ratio (GCr) is implemented to assess and flag recordings above a predetermined threshold.
        The GCr increases non-linearly with increasing stimulation voltage (Figure \ref{fig:PSD_correct}c) and demonstrated significant differences between impedance mismatches for a fixed stimulation voltage (Figure \ref{fig:gcr}).
        
        We test the proposed preprocessing strategy \textit{in vitro}, expecting to see all oscillatory power calculations converge to the \SI{0}{\volt} calculation.
        The preprocessing removed broadband differences between PSDs collected at different voltages (Figure \ref{fig:PSD_correct}a,c).
        Oscillatory power calculation in all bands demonstrated a normalization to the \SI{0}{\volt} measurement across all stimulation amplitudes (Figure \ref{fig:PSD_correct}b,d).
        Residual differences in $\theta$ are evident (Figure \ref{fig:PSD_correct}d), but smaller than the uncorrected PSDs (Figure \ref{fig:PSD_correct}b).
       
\section{Discussion}
    A new class of bidirectional DBS devices capable of \textit{differential} LFP ($\partial$LFP) channels enable reliable recordings alongside clinical DBS therapy, even in the presence of active stimulation, by removing signals seen equally in the two recording electrode\cite{gilron2021,starr2018totally,stanslaski2012,stanslaski2018,swann2017chronic,cummins2021chronic}.
    However, dynamics in the surrounding tissue can change the impedance mismatch between recording electrodes and confound neural oscillations with impedance-related gain compression, a process we characterized here as mismatch compression (MC).
    Next-generation devices are being engineered to better account for these, and other, artifacts in hardware \cite{stanslaski2018,anso2022concurrent}.
    However, an analytical approach to mismatch compression is needed to reliably predict artifacts, adjust analyses to minimize inclusion of severely distorted LFP arising from impedance mismatch changes, and to maximize the signal in precious datasets.
    an analytical approach to mitigating the effects of MC is needed to better analyse already collected datasets and to engineer low-power bdDBS devices for chronic aDBS.
    
    In this study, we took a multimodal approach, spanning \textit{in vivo}, \textit{in silico}, and \textit{in vitro} recordings, to characterize and mitigate the effects of MC for more reliable oscillatory readout analysis.
    We first observed significant changes in the recordings and in the recording environments (Figure \ref{fig:Z_diff}).
    We then showed that a reduced \textit{in silico} model could regenerate these observations in simulation, validating the emergence of MC distortions \textit{in vitro} (Figure \ref{fig:stim_arts}).
    Finally, we used the \textit{in silico} model to develop a mitigation strategy for $\partial$LFP MC distortions (Figure \ref{fig:mc_mitigation}) that worked to normalize oscillatory power calculations \textit{in vitro} (Figure \ref{fig:PSD_correct}).
        
    \subsection{Recording Oscillations with $\partial$LFP}
        Oscillations in LFPs reflect synaptic inputs into gray matter and correlate with function \cite{buzsaki2012,little2014functional}.
        Efforts to integrate oscillations into adaptive DBS applications have demonstrated early success in the relatively uniform targets used in Parkinson's Disorder \cite{starr2018totally,swann2017chronic,staub2016ep,tan2019decoding,swann2018adaptive}.
        However, adaptive DBS requires readouts that are accurate over long periods of time, in the presence of numerous sources of variability and confounds, including the DBS artifact itself.
        
        When amplifiers receive an input larger than they are designed to properly gain, the resulting output is \textit{compressed} compared to the desired output, a process called \textit{gain compression} (Figure \ref{fig:model_overview}d) \cite{schubert2015fundamentals}.
        $\partial$LFP recordings rely on recording from two electrodes around the stimulation electrode (Figure \ref{fig:dLFP_overview}a), subtracting signals they share in common (common-mode rejection), and amplifying signals that are different (differential-mode) \cite{stanslaski2012}.
        However, if the DBS artifact is not seen equally in both electrodes, a small fraction of residual artifact can overwhelm a signal amplifier tuned for much smaller neural oscillations (Figure \ref{fig:model_overview}c).
        Importantly, this distortion can be subtle through a \textit{soft-clipping} that is difficult to visually identify, with overt \textit{hard-clipping} an extreme case (Figure \ref{fig:model_overview}d).

    \subsection{Impedances and Mismatch Compression}
        DBS electrode impedances affects the amplitude of stimulation delivered to the target \cite{butson2005tissue,wong2018,sillay2010,satzer2014variation} and can attenuate field potential recordings \cite{carron2019}.
        Impedance changes over weeks and months, potentially introducing confounding distortions to oscillatory analyses at the scale of disease recovery \cite{wong2018,sillay2010,satzer2014,swann2017chronic,miocinovic2009experimental}.
        Impedances in the $\partial$LFP channel can be highly variable due to heterogeneity in surrounding brain tissue \cite{logothetis2007vivo}.
        A growing class of white matter structures targeted in \textit{connectomic DBS} are particularly susceptible to impedance mismatches due to surrounding tissue heterogeneity \cite{logothetis2007vivo,veerakumar2019field,latikka2001, horn2020opportunities, haber2019prefrontal, howell2019quantifying}.
        $\partial$LFP channels assume artifact is recorded equally, but electrode impedances can attenuate artifacts differently in the two $\partial$LFP electrodes \cite{miocinovic2009experimental,butson2005tissue,wong2018,sillay2010,satzer2014variation,logothetis2007vivo}.
        When this impedance mismatch occurs it can lead to incomplete artifact rejection and subsequent gain compression in recordings, a process we introduce and characterize as \textit{mismatch compression} (MC).
        
    \subsection{Impedance mismatches in SCCwm-DBS}
        The SCCwm target has been well studied for antidepressant applications \cite{mayberg2005,riva2018connectomic,holtzheimer2017} and is a white matter target with gray matter around it \cite{veerakumar2019field}.
        We observed significant impedance mismatches in our six SCCwm-DBS patients, and this mismatch is dynamic over the first seven months of therapy, when initial antidepressant effects are present alongside increased emotional reactivity \cite{crowell2015} (Figure \ref{fig:Z_diff}).
        These mismatches are accompanied by variability in the ORM power, a constant amplitude signal that appears to change in the presence of amplifier saturation (Figure \ref{fig:in_vivos}a).
        Together with the expected failure mode of the PC+S\texttrademark\space in the presence of impedance differences, this suggests impedance mismatches must be properly accounted for in oscillatory recordings alongside SCCwm-DBS in order to reliably identify electrophysiologic correlates of disease without conflating impedance mismatches.
        
    \subsection{Mismatch Compression distortions}
        A generalized model of MC was developed to explain observed empirical changes (Figure \ref{fig:stim_arts}).
        Simulated $\partial$LFP grossly exhibited the frequency-domain changes (Figure \ref{fig:stim_arts}a,c) seen in empirical recordings (Figure \ref{fig:in_vivos}).
        Simulated PSDs exhibited slope flattening and emergence of narrow-band peaks that are vivid in simulated impedance mismatch (Figure \ref{fig:stim_arts}b,d).
        The emergence of intermodulation harmonics (IMH) and flattening of the PSD with stimulation voltage supported MC as a parsimonious explanation of empirical observations (Figure \ref{fig:stim_arts}e).
    
        \textit{In vitro} recordings were taken in variable impedance mismatch and demonstrated distortions predicted by \textit{in silico} MC model (Figure \ref{fig:stim_arts}f).
        The distortions themselves were stereotyped: broad-band flattening of PSD slope and narrow-band \textit{intermodulation harmonics} (IMH) at fixed frequencies.
        IMH, in addition to the stimulation shaping harmonics (SSH) and the aliased stimulation harmonics (ASH), resulted in a distinct pattern of narrowband peaks when stimulating at \SI{130}{\hertz} and sampling at \SI{422}{\hertz}: largest at \SI{32}{\hertz}, \SI{64}{\hertz}, and \SI{66}{\hertz}.
        
        Other peaks were apparent on visual inspection of \textit{in vitro} recordings under all conditions, including \SI{0}{\volt} and are considered device-specific artifacts not related to mismatch compression.
        Recordings taken in uniform media (Figure \ref{fig:agar_exp}a) were compared to those taken at the agar-saline interface with mismatched resistivities (Figure \ref{fig:agar_exp}b).
        Uniform media recordings demonstrated a large stimulation-dependent peak centered at \SI{32}{\hertz} that is absent at \SI{0}{\volt} stimulation.
        
    \subsection{Preprocessing removes distortable features}
        Recent focus on properly decomposing LFPs into oscillatory and non-oscillatory components allow us to better extract neural signals \cite{donoghue2020}, but large stimulation artifacts can make this challenging \cite{kent2015}.
        Minimizing the effect of MC distortion is crucial for a reliable long-term disease readout in $\partial$LFP recordings.
        Completely removing MC distortions \textit{post hoc} is likely impossible given the noninvertibility of gain compression; hardware improvements are necessary.
        However, steps to mitigate the effects of MC in the meantime are possible by removing features that can be distorted.
        
        We proposed and validated a mitigation pipeline that removes frequency-domain features susceptible to MC (Figure \ref{fig:mc_mitigation}).
        \begin{table*}
    		\centering
    		\begin{tabular}{c c c c c c} 
    			& $\delta$ & $\theta$ & $\alpha$ & $\beta$ & $\gamma$ \\ \hline
    			Standard & \SIrange{1}{4}{\hertz} & \SIrange{4}{8}{\hertz} & \SIrange{8}{14}{\hertz} & \SIrange{14}{30}{\hertz} & \SIrange{30}{50}{\hertz} \\ \hline
    			Adjusted & \SIrange{1}{4}{\hertz} & \SIrange{4}{8}{\hertz} & \SIrange{8}{14}{\hertz} & \SIrange{14}{20}{\hertz} & \SIrange{40}{50}{\hertz}
    		\end{tabular}
    		\caption{Oscillatory Bands in $\partial$LFP. Traditionally defined bands, labeled in their greek letters. Adjusted bands for dissertation analysis.}
    		\label{tab:prelim_oscbands}
    	\end{table*}
    	The major steps of the pipeline involve polynomial fit of the full PSD, followed by adjusted oscillatory band ranges to avoid MC related narrow-band artifacts (Table \ref{tab:prelim_oscbands}).
        The pipeline was validated across a typical therapeutic ranges \SIrange{3.5}{5}{\volt} in \textit{in vitro} and \textit{in silico} models.
        The mitigation pipeline brings all power calculations closer to to each other, and to the noise floor reflecting ground truth.
        Residual variability in $\theta$ is evident, suggesting residual MC-related artifacts, but the variability is much smaller than the variability without mitigation and in the reverse direction (Figure \ref{fig:PSD_correct}b vs Figure \ref{fig:PSD_correct}d).
        As a final step, the GCr can be calculated and used as a final check for principled recording exclusion based on an \textit{a priori} threshold.
        
        This preprocessing strategy takes a cautious approach and removes features entirely; it likely removes meaningful neural signals, like $\frac{1}{f}$ slope \cite{veerakumar2019field,donoghue2020}, but this enables more confidence in oscillatory power that survives correction.
        While the latest generation of $\partial$LFP bdDBS devices have improved signal and noise profiles, the MC process itself is intrinsic to $\partial$LFP hardware.
        Care must be taken to rule out MC in any readout applications, acute or chronic, if using a $\partial$LFP recording channel \cite{cummins2021chronic,anso2022concurrent}.
        Additionally, this mitigation strategy may serve lasting utility as a signal processing step for low-cost bdDBS devices using $\partial$LFP channels.
        
    \subsection{Limitations} 
        This work has several limitations.
        First, the reduced $\partial$LFP model is developed in a general way to focus on gain compression secondary to impedance mismatches, and not to recapitulate all device-specific artifacts \cite{swann2017chronic}.
        Second, the tissue-electrode interface ignores capacitance in impedance \cite{lempka2009} and does not account for distortions in phase spectrum, only power spectral analyses.
        Third, the preprocessing strategy avoids features that can be distorted by MC, it doesn't directly invert the MC process, resulting in almost certain loss of information in broad slope signals and certain oscillatory subranges \cite{veerakumar2019field}.
        Further, the restriction on oscillatory band ranges has implications for the generalizability of results to the broader ranges defined in the literature.
        Finally, we study and model MC in the presence of DBS artifact but other large-amplitude artifacts can drive MC, such as ECG and EMG signals evident in the PC+S\texttrademark \cite{swann2017chronic}.
        While the contemporary generation of devices have reduced sources of noise in recording, MC must still be accounted for in any $\partial$LFP channel \cite{cummins2021chronic,anso2022concurrent,stanslaski2018}.
            
\section{Conclusion}
    While differential LFP goes a long way towards removing the DBS waveform, heterogeneity and electrical property dynamics in the surrounding brain target can distort oscillatory power measurements, a process we characterize and mitigate as \textit{mismatch compression}.
    These distortions must be accounted for and/or corrected; our work here provides an important correction pipeline for reliable, long-term disease readouts in $\partial$LFP oscillations.
    Further work to more explicitly invert the MC distortions using the models developed here could enable low-cost bdDBS devices and inform design specifications tailored to connectomics DBS applications.

\section*{Conflicts of Interest}
    CCM is a paid consultant for Boston Scientific Neuromodulation, receives royalties from Hologram Consultants, Neuros Medical, Qr8 Health, and is a shareholder in the following companies:  Hologram Consultants, Surgical Information Sciences, CereGate, Autonomic Technologies, Cardionomic, Enspire DBS.
    HM has a consulting agreement with Abbott Labs (previously St Jude Medical, Neuromodulation), which has licensed her intellectual property to develop SCC DBS for the treatment of severe depression (US 2005/0033379A1). 
    RG serves as a consultant to and receives research support from Medtronic, and serves as a consultant to Abbott Labs. 
    The terms of these arrangements have been approved by Emory University, Icahn School of Medicine, and Duke University in accordance with policies to manage conflict of interest. 
    All other authors have no COI to declare.

\section*{Acknowledgment}
    Thanks to Scott Stanslaski at Medtronic for technical assistance with the Activa PC+S\texttrademark\space device. 
    Thank you to Dr. Riley Zeller-Townshend, Nathan Kirkpatrick, Patrick Howard, and Mosadoluwa Obatusin for feedback throughout.
    Thank you to Dr. Warren Grill for discussions and guidance in study design.
    Thank you to the clinical team, particularly Sinead Quinn, and Lydia Denison. 

\section*{Funding Sources}
NIH R01 MH106173, NIH BRAIN UH3NS103550, and the Hope for Depression Research Foundation and  The DBS Activa PC+S research devices were donated by Medtronic (Minneapolis, MN).

\ifCLASSOPTIONcaptionsoff
  \newpage
\fi

\bibliographystyle{IEEEtran}
\bibliography{refs.bib}

\end{document}